\documentclass{article}
\usepackage{amsfonts}

\usepackage{amssymb}
\usepackage{graphicx}
\usepackage{amsmath}

\numberwithin{equation}{section}

\newcommand{\la}{\langle}
\newcommand{\ra}{\rangle}

\input{tcilatex}

\begin{document}

\title{The Sufficient Optimality Condition for Quantum Information Processing%
\thanks{%
Originally published in: Radio Eng. Electron. Phys., \textbf{19} (7),
p.\thinspace 39, 1974. [trans. from Radiotekhnika i Electronika, 1974, 
\textbf{19}, 7, 1391--1395]}}
\author{V\ P\ Belavkin \\
School of Mathematical Sciences, University of Nottingham, \\
Nottingham NG7 2RD \\
E-mail\textup{:} vpb@maths.nott.ac.uk \\
A\ G\ Vantsyan}
\maketitle

\begin{abstract}
The necessary and sufficient conditions of optimality of the decoding of
quantum signals minimizing the Bayesian risk are generalized for the Shannon
mutual information criteria. It is shown that for a linear channel with
Gaussian boson noise these conditions are satisfied by coherent
quasi-measurement of the canonical annihilation amplitudes in the received
superposition. 
\end{abstract}

\section{Necessary and sufficient conditions of optimality}

In~\cite{1a,2a} dealing with optimization of the reception of quantum
signals such as electromagnetic waves in the optical band, the search for
the necessary conditions of optimality in the class of randomized strategies
based on indirect measurements was our main concern. According to this
universal approach, we shall specify the randomized strategies by the
operator probability measures $\Pi (\mathrm{d}\beta )$ corresponding to
quasi-measurements of certain noncommuting observables $b_{j}=\int \beta
_{j}\Pi (\mathrm{d}\beta )$ such that 
\begin{equation*}
\Pi (\mathrm{d}\beta )\geq 0,\;\;\int \Pi (\mathrm{d}\beta )=\hat{1}.
\end{equation*}%
The equations derived in~\cite{1a,2a} have the same form for both risk and
information criteria of optimality: 
\begin{equation}
\left( R(\beta )-\Lambda \right) \Pi (\mathrm{d}\beta )=0,\quad \Lambda
=\int R(\beta )\Pi (\mathrm{d}\beta ),  \label{one a}
\end{equation}%
where $R(\beta )=\int c(\vartheta ,\beta )\rho (\vartheta )P(\mathrm{d}%
\vartheta )$ is the "\textit{a posteriori}" risk or nega-information
operator in the Hilbert space of quantum states $\mathbb{H}$. Here $\rho
(\vartheta )$ is the family of density operators describing the state of the
quantum channel depending on the transmitted information $\vartheta $ with a
prior distribution $P(\mathrm{d}\vartheta )$ and $c(\beta ,\vartheta )$ is a
given penalty function in Bayes case and the random information 
\begin{equation}
i(\vartheta ,\beta )=\ln \frac{P(\mathrm{d}\beta |\vartheta )}{\int P(%
\mathrm{d}\beta |\vartheta )P(\mathrm{d}\vartheta )},\quad P(\mathrm{d}\beta
|\vartheta )=\mathrm{Tr}\Pi (\mathrm{d}\beta )\rho (\vartheta )
\label{two a}
\end{equation}
with the opposite sign, $c\left( \vartheta ,\beta \right) =-i\left(
\vartheta ,\beta \right) $, for the optimization criterion of maximum
Shannon information 
\begin{equation*}
\mathsf{I}_{\beta ,\vartheta }=\iint \ln \frac{P(\mathrm{d}\beta |\vartheta )%
}{P(\mathrm{d}\beta |\vartheta )P(\mathrm{d}\vartheta )}P(\mathrm{d}\beta
|\vartheta )P(\mathrm{d}\vartheta ).
\end{equation*}%
It is obvious that operators $\Pi ^{\mathrm{o}}(\mathrm{d}\beta )$
satisfying equation~\eqref{one a} are degenerate (provided $R(\beta
)-\Lambda \neq 0$) and for each $\beta $ have a range of values belonging to
the zero eigensubspace of the difference $R(\beta )-\Lambda $. If operators $%
B(\beta )\equiv R(\beta )-\Lambda $ have a unique eigenvector $\varphi
_{\beta }$ for each $\beta $, corresponding to the zero eigenvalue, then the
operator measure $\Pi ^{\mathrm{o}}(\mathrm{d}\beta )$ is proportional to
the projection operators $\Pi ^{\mathrm{o}}(\mathrm{d}\beta )=\varphi
_{\beta }^{\mathrm{o}}\varphi _{\beta }^{\mathrm{o}\ast }\mathrm{d}\beta $.
In the general case when degeneration of the zero eigenvalue of operators $%
B(\beta )$ is possible: $B(\beta )\varphi _{\beta \nu }=0$, $\nu \in N(\beta
)$, each resolution of identity $\int \Pi (\mathrm{d}\beta )=\hat{1}$
satisfying Equation.~\eqref{one a} may be included in some more detailed
resolution $\int \varphi _{\gamma }\varphi _{\gamma }^{\ast }\mathrm{d}%
\gamma =\hat{1}$ for $\gamma =(\beta ,\nu )$. For different $\gamma $ the
vectors $\varphi _{\gamma }$ describing the \textquotedblleft
elementary\textquotedblright\ measurements need not necessarily be
orthogonal: $\varphi _{\gamma }^{\ast }\varphi _{\gamma ^{\ast }}\neq 0$.

In Bayes case the sufficient conditions for optimality are very simple: the
operators $\Pi ^{\mathrm{o}}(\mathrm{d}\beta )$ satisfying Equation.~%
\eqref{one a} minimize the average risk 
\begin{equation}
\mathsf{R}=\big\la c(\vartheta ,\beta )\big\ra=\mathrm{Tr}\int R(\beta )\Pi (%
\mathrm{d}\beta )  \label{three a}
\end{equation}%
if and only if the condition of nonnegative definiteness 
\begin{equation}
B(\beta )\equiv R(\beta )-\int R(\beta ^{\prime })\Pi ^{\mathrm{o}}(\mathrm{d%
}\beta ^{\prime })\geq 0  \label{four a}
\end{equation}%
is satisfied for all $\beta $. Actually, for any other operators measure $%
\Pi (\mathrm{d}\beta )\neq \Pi ^{\mathrm{o}}(\mathrm{d}\beta )$ the
difference%
\begin{equation*}
\mathsf{R}-\mathsf{R}^{\mathrm{o}}=\mathrm{Tr}\left[ \int R(\beta )\Pi (%
\mathrm{d}\beta )-\int R(\beta )\Pi ^{\mathrm{o}}(\mathrm{d}\beta )\right] =%
\mathrm{Tr}\int B(\beta )\Pi (\mathrm{d}\beta )
\end{equation*}
is nonnegative since it is the trace of a sum of products of nonnegative
operators $B(\beta ),\Pi (\mathrm{d}\beta )$.

Conditions~\eqref{one a} and~\eqref{four a} are applicable also for the
optimization of the processing of quantum signals according to the maximum
likelihood criterion. For this it is sufficient to consider that this
criterion can be formally taken as Bayes criterion with uniform
(unnormalized) \textit{a priori} distribution $P(\mathrm{d}\vartheta )=%
\mathrm{d}\vartheta $ and a simple penalty function $c(\beta ,\vartheta
)=-\delta (\vartheta -\beta )$. This means that the a posteriori risk
operator $R(\beta )$ should in this case be replaced by the density operator 
$\rho (\vartheta )$ at the estimate point $\vartheta =\beta $. We shall call
the quantum strategies $\Pi ^{\mathrm{o}}(\mathrm{d}\beta )$ satisfying
conditions~\eqref{one a} and~\eqref{four a} for $R(\beta )=-\rho (\beta )$
optimum with respect to the maximum likelihood criterion. We shall give the
solution of the problem of the discrimination of nonorthogonal signals for
the following simplest case.

In~\cite{3a} the concept of coherent processing a boson\footnote{%
We recall that we are giving the name boson signal to the quantum signal
described by the operators $\{\alpha _{\nu },\alpha _{\nu }^{*}\}$
satisfying the commutation relations $\alpha _{\nu }\alpha _{\nu ^{\prime
}}-\alpha _{\nu ^{\prime }}\alpha _{\nu }=0,\,\alpha _{\nu }\alpha _{\nu
^{\prime }}^{*}-\alpha _{\nu ^{\prime }}^{*}\alpha _{\nu }=\delta _{\nu \nu
^{\prime }}$. In particular, the optical signal is described by the photon
annihilation and creation operators $a$ and $a^{*}$ respectively.} signal $%
b=(b_{\nu })$, i.e., of indirect linear measurement realized by measuring
the superposition $b+a_{0}^{\ast }$, where $a_{0}$ is vacuum boson noise,
was introduced. The question of the physical realization of this measurement
was discussed at the Third All-Union Conference on the Physical Principles
of Information Transmission by Laser Radiation. In particular it was shown~%
\cite{4a} that coherent measurement of a narrowband optical signal can be
realized by using an ideal heterodyne reception (ideal count of photons at
different points of superposition of the received and the reference waves).
The backward vacuum wave radiated by an ideally matched receiver into the
communication line plays the role of noise $a_{0}$.

In~\cite{3a} the quality of such processing was also defined from the
maximum likelihood criterion for the case where $b$ is the superposition of
a coherent signal $\vartheta =(\vartheta _{\nu })$ and a Gaussian boson
noise $a$. The use of equations~\eqref{one a} and~\eqref{four a} and a
suitable representation of the density operator makes it obvious that the
processing described by the coherent projectors 
\begin{equation}
\Pi (\mathrm{d}\beta )=|\beta \rangle \,\langle \beta |\mathrm{d}\mu (\beta
),\quad \mathrm{d}\mu (\beta )=\prod_{\nu }\tfrac{1}{\pi }\mathrm{d}\func{Re}%
\beta _{\nu }\mathrm{d}\func{Im}\beta _{\nu }  \label{five a}
\end{equation}%
is optimal. Such a suitable representation of the density operator $\rho
(\vartheta )$ of the displaced Gaussian state $b=\vartheta +a$ is the
representation in the form of the expression 
\begin{equation}
\rho (\vartheta )=|L|^{-1}:\exp \big\{b-\vartheta )^{\dagger
}L^{-1}(b-\vartheta )\big\}:  \label{six a}
\end{equation}%
normally ordered with respect to the operators $b^{\ast },b$. Here $L\big\|%
\langle \alpha _{\nu }\alpha _{\nu }^{\ast }\rangle \big\|$ is the
correlation matrix of the noise $a$, the colon-brackets $:\cdot :$ denote
normal order such that the operators $a^{\ast }$ act to the left after the
operators $a$, and $|L|\equiv \det L$.

Putting $R(\beta )=-\rho (\beta ),\Lambda =-|L^{-1}|\hat{1}$, and
considering the well-known~\cite{5a} properties $\int |\beta \rangle
\,\langle \beta |\mathrm{d}\mu (\beta )=1$,%
\begin{equation*}
:p(b^{\ast },b):|\beta \rangle =p(b^{\ast },\beta )\,|\beta \rangle 
\end{equation*}%
of the coherent vectors, we at once find that equation~\eqref{one a} has a
unique solution coinciding with~\eqref{five a}. The nonnegative definiteness
of the operator 
\begin{equation*}
B(\beta )=|L^{-1}|\Big(1-:\exp \big\{-(b-\vartheta )^{\dagger
}L^{-1}(b-\vartheta )\big\}:\Big)
\end{equation*}%
is beyond doubt.

If the signals $(\vartheta _{t})$ are known apart from the phase, the
coherent processing is no longer optimum; however, it remains quasioptimal
if the dimensionality of $\vartheta $ is large.

\section{Local optimality according to information criterion}

In the case of the information criterion it is difficult to get a global
criterion of optimality of the solutions $\Pi ^{\mathrm{o}}(\mathrm{d}\beta )
$ of equation~\eqref{one a} minimizing the \textquotedblleft
Shannon\textquotedblright\ risk $\mathsf{R}=-\mathsf{I}_{\beta ,\vartheta }$
in view of its nonlinear dependence on $\Pi (\mathrm{d}\beta )$ [the
\textquotedblleft penalty function\textquotedblright ~\cite{2a} also depends
on $\Pi (\mathrm{d}\beta )$]. Therefore we give a differential criterion of
optimality.

We shall restrict the discussion to operator probabilities of degenerate
form $\Pi (\mathrm{d}\beta )=\varphi _{\beta }\varphi _{\beta }^{\ast }%
\mathrm{d}\beta $, where $(\varphi _{\beta })$ is the complete family of
vectors $\int \varphi _{\beta }\varphi _{\beta }^{\ast }\mathrm{d}\beta =%
\hat{1}$. It can be shown that this is sufficient for the verification of
local optimality $\delta ^{2}\mathsf{R}>0$ of the degenerate solutions $\Pi
^{\mathrm{o}}(\mathrm{d}\beta )=\varphi _{\beta }^{\mathrm{o}}\varphi
_{\beta }^{\mathrm{o}{\normalsize \ast }}\mathrm{d}\beta $ of Equation.~%
\eqref{one a}.

Making use of the dependence of the variations $\delta \varphi _{\beta
}=\varphi _{\beta }-\varphi _{\beta }^{\mathrm{o}}$ 
\begin{equation*}
\int (\varphi _{\beta }^{\mathrm{o}}\delta \varphi _{\beta }^{\ast }+\delta
\varphi _{\beta }\varphi _{\beta }^{\mathrm{o}\ast }+\delta \varphi _{\beta
}\delta \varphi _{\beta }^{\ast })\,\mathrm{d}\beta =0,
\end{equation*}%
it is not difficult to find the increment $\Delta \mathsf{R}=\mathsf{R}-%
\mathsf{R}^{\mathrm{o}}$ of Shannon risk~\eqref{three a} (with the penalty
function~\eqref{two a} depending on $\varphi _{\beta }$) at the stationary
point $\varphi _{\beta }^{\mathrm{o}}$ with an accuracy up to second-order
terms in $\delta \varphi _{\beta }$: 
\begin{align}
& \Delta \mathsf{R}\simeq \iint \bigg\{c\left( \vartheta ,\beta \right)
\delta p(\beta |\vartheta )P(\mathrm{d}\vartheta )  \notag  \label{seven a}
\\
& =\tfrac{1}{2}\left[ \big(\delta \ln p(\beta |\vartheta )\big)^{2}-\big(%
\delta \ln p(\beta )\big)^{2}\right] p(\beta |\vartheta )p(\mathrm{d}%
\vartheta )\bigg\}  \notag \\
& =\int \bigg\{\delta \varphi _{\beta }^{\ast }B(\beta )\delta \varphi
_{\beta }=\tfrac{1}{2}\Big[\int (\delta \varphi _{\beta }^{\ast }\psi
_{\beta }(\vartheta )+\psi _{\beta }^{\ast }(\vartheta )\delta \varphi
_{\beta })^{2}p(\beta |\vartheta )P(\mathrm{d}\vartheta )  \notag \\
& -(\delta \varphi _{\beta }^{\ast }\psi _{\beta }+\psi _{\beta }^{\ast
}\delta \varphi _{\beta })^{2}p(\beta )\Big]\bigg\}\,\mathrm{d}\beta ,
\end{align}%
where $B(\beta )$ is the difference~\eqref{four a}, 
\begin{equation*}
\psi _{\beta }(\vartheta )=\frac{\rho (\vartheta )\varphi _{\beta }^{\mathrm{%
o}}}{p(\beta |\vartheta )},\quad p(\beta |\vartheta )=\varphi _{\beta }^{%
\mathrm{o}\ast }\rho (\vartheta )\varphi _{\beta }^{\mathrm{o}},
\end{equation*}%
and the vector $\psi _{\beta }=\rho \varphi _{\beta }^{\mathrm{o}}\big/%
p(\beta )$ ($p(\beta )=\int p(\beta |\vartheta )P(\mathrm{d}\vartheta )$) is
the vector $\psi _{\beta }(\vartheta )$ averaged with the Bayessian
posterior density%
\begin{equation*}
p(\vartheta |\beta )=p(\beta |\vartheta )p(\vartheta )\big/\int p(\beta
|\vartheta )P(\mathrm{d}\vartheta ).
\end{equation*}

A simple analysis of the positiveness $\delta ^{2}\mathsf{R}>0$ of the
variation~\eqref{seven a} of Shannon risk shows that in contrast to the
Bayes case the nonnegativeness of the operators $B(\beta )\geq 0$ is
necessary, but not sufficient for the local optimality of the solutions $%
\varphi _{\beta }^{\mathrm{o}}$ of the equation $B(\beta )\varphi _{\beta }=0
$: the additional term in~\eqref{seven a} (in square brackets) has the
meaning of a posteriori variance of the real random quantity $2\func{Re}%
\delta \varphi _{\beta }^{\ast }\psi _{\beta }(\vartheta )$ and is generally
positive.

Let, for example, the density operator have the form~\eqref{six a} and the
prior distribution $P(d\vartheta )$ be Gaussian in the multidimensional
space of the information parameters $\vartheta =(\vartheta _{\nu })$: 
\begin{equation*}
P(d\vartheta )=|S|^{-1}\exp \{-\vartheta ^{\dagger }S^{-1}\vartheta \}\,d\mu
(\vartheta ),\quad d\mu (\vartheta )=\prod_{\nu }\tfrac{1}{\pi }d\func{Re}%
\vartheta _{\nu }d\func{Im}\vartheta _{\nu }.
\end{equation*}

We shall check the local optimality of the coherent solutions~\eqref{five a}
of equation~\eqref{one a} in the boson Gaussian case according to the
information criterion. As shown in~\cite{2a}, the coherent vectors~%
\eqref{five a} satisfy Equation.~\eqref{one a} and the operator $B(\beta )$
has a quadratic Gaussian form: 
\begin{equation*}
B(\beta )=(b-\beta )^{\dagger }H\rho (b-\beta ),\,\rho
=|L+S|^{-1};e^{b^{\dagger }(L+S)^{-1}b}:,
\end{equation*}%
where the matrix $H=L^{-1}-(S+L)^{-1}$ is not larger than unity in
accordance with the inequalities $S\geq 0,L\geq 1,0\leq H\leq 1$.
Considering the analytic dependence of the function 
\begin{equation*}
\psi _{\beta }(\vartheta )=\frac{\rho (\vartheta )|\beta \rangle }{\langle
\beta |\rho (\vartheta )|\beta \rangle }=e^{-(b-\beta )^{\dagger
}L^{-1}(\beta -\vartheta )}|\beta \rangle 
\end{equation*}%
on $\vartheta $ (i.e., the independence on $\vartheta ^{\ast }$) and
carrying out conditional averaging over $\vartheta $ in~\eqref{seven a},%
\begin{equation*}
\int \big(\delta \varphi _{\beta }^{\ast }\psi _{\beta }(\vartheta )+\psi
_{\beta }^{\ast }(\vartheta )\delta \varphi _{\beta }\big)^{2}p(\vartheta
|\beta )d\mu (\vartheta )
\end{equation*}
with the density%
\begin{equation*}
p(\vartheta |\beta )=|M|\exp \big\{-(\vartheta -A\beta )^{\dagger
}M(\vartheta -A\beta )\big\},
\end{equation*}%
where $\,A=S(S+L^{-1},\,M=S^{-1}+L^{-1}$, we find that in the Gaussian case
the variation $\delta ^{2}\mathsf{R}$ has the form 
\begin{align*}
\delta ^{2}\mathsf{R}& =\int \delta \varphi _{\beta }^{\ast }\big(B(\beta
)-D(\beta )\big)\delta \varphi _{\beta }\,d\mu (\beta ), \\
d\mu (\beta )& =\prod_{\nu }\tfrac{1}{\pi }d\func{Re}\beta _{\nu }d\func{Im}%
\beta _{\nu },
\end{align*}%
where 
\begin{align*}
D(\beta )& =:p(b)\left[ e^{-(b-\beta )^{\dagger }(1-H)(b-\beta
)}-e^{-(b-\beta )^{\ast }(b-\beta )}\right] :\geq 0, \\
p(b)& =|S+L|^{-1}\exp \big\{-b^{\dagger }(S+L)^{-1}b\big\}.
\end{align*}

Thus, in order to prove the optimality of coherent quasi-measurement for the
information criterion one should verify the operator inequality $B(\beta
)-D(\beta )\geq 0$ or 
\begin{align}
& p(\beta )e^{-b^{\dagger }(S+L)^{-1}\beta }:\Big[b^{\dagger
}He^{-b^{\dagger }(S+L)^{-1}b}b  \notag  \label{eight a} \\
& -\left( e^{-b^{\dagger }(H-1)b}-e^{-b^{\dagger }b}\right) \Big]:e^{-\beta
^{\dagger }(S+L)^{-1}b}\geq 0
\end{align}%
(here the change of variables $b-\beta \rightarrow b$ has been carried out).
The operator occurring on the left-hand side of~\eqref{eight a} has the
structure $A_{\beta }^{\ast }:[\cdot ]:A_{\beta }$ and is positive only if
the operator in the square brackets is positive. Considering that the
operator inequality%
\begin{equation*}
:e^{-b^{\dagger }(S+L)^{-1}b}:\geq :e^{-b^{\dagger }(1-H)b}:
\end{equation*}
is satisfied by virtue of the matrix inequality $(S+L)^{-1}=L^{-1}H\leq 1-H$%
, we find that inequality~\eqref{eight a} is satisfied if 
\begin{equation}
:b^{\dagger }He^{b^{\dagger }(H-1)b}b:\geq :e^{b^{\dagger
}(H-1)b}-e^{-b^{\dagger }b}:.  \label{nine a}
\end{equation}%
This inequality~\eqref{nine a} becomes obvious in the diagonal
representation in the occupation numbers $n_{\nu }$: 
\begin{equation*}
\prod_{\nu }h_{\nu }^{n_{\nu }}\sum_{\nu }n_{\nu }\geq \prod_{\nu }h_{\nu
}^{n_{\nu }}\quad \text{for}\quad \sum_{\nu }n_{\nu }\neq 0,
\end{equation*}%
where $h_{\nu }$ are the eigenvalues of the matrix $H=L^{-1}-(S+L)^{-1}$.
For $\sum_{\nu }n_{\nu }=0$ both the left- and right-hand sides of
inequality~\eqref{nine a} vanish.


\begin{thebibliography}{9}
\bibitem{3a} V.P. Belavkin, \textit{Radiotekhnika i Elektronika}, 1972, 
\textbf{17}, No.\thinspace 12, 2533 [Radio Eng. Electron. Phys., \textbf{17}
No.\thinspace 12 (1972)].

\bibitem{4a} V.P. Belavkin, Coherent measurement of optical signals, Proc.
of the Third All-Union Conference on the Physical Principles of Recording
and Processing Inofrmation by Laser Radiation, Kiev, 1973, I, p.\,7.

\bibitem{2a} V.P. Belavkin and R.L. Stratonovich, \textit{Radiotekhnika i
Elektronika,} 1973, \textbf{19}, 9, 1839 [Radio Eng. Electron. Phys., 
\textbf{19} 9 (1973)].

\bibitem{5a} J. Klauder and E. Sudershan, Fundamentals of Quantum Optics (in
Russian transl.), Mir Press, 1970.

\bibitem{1a} R.L. Stratonovich, The quantum generalization of optimal
statistical estimation and hypothesis testing, \textit{J. of Stochastics},
1973, \textbf{1}, 87--126.
\end{thebibliography}
\end{document}